\documentclass[a4paper, amsfonts, amssymb, amsmath, reprint, showkeys, nofootinbib, twoside]{revtex4-2}
\usepackage[english]{babel}
\usepackage[utf8]{inputenc}
\usepackage[colorinlistoftodos, color=green!40, prependcaption]{todonotes}
\usepackage{amsthm}
\usepackage{mathtools}
\usepackage{physics}
\usepackage{xcolor}
\usepackage{graphicx}
\usepackage[left=23mm,right=13mm,top=35mm,columnsep=15pt]{geometry} 
\usepackage{adjustbox}
\usepackage{placeins}
\usepackage[T1]{fontenc}
\usepackage{lipsum}
\usepackage{csquotes}
\usepackage{natbib}
\usepackage[title]{appendix}
\graphicspath{ {images/} } 
\usepackage[pdftex, pdftitle={Article}, pdfauthor={Author}]{hyperref} 
\hypersetup{
colorlinks=true,
linkcolor=blue,
urlcolor=cyan,
}

\begin{document}
\title{A no-go theorem for Quantum theory ontological models}

\author{Tung Ten Yong}
    \email[Correspondence email address: ]{tytung2020@gmail.com}

\date{\today} 

\begin{abstract}
In this paper, we show that Quantum Mechanics does not admit ontological models, in the sense that the quantum state of a system cannot correspond to a set of physical states representing the independent reality of the system. We show, via two thought experiments based on the Wigner’s friend scenario, that if the ontic state of physical systems in the lab is the same for Wigner and for his friend, one of the following will be violated: PBR theorem, quantum-theoretic predictions and the "No-superdeterminism" assumption.
\end{abstract}


\maketitle

\section{Outline}

Since the early days of Quantum Mechanics, there has been unending discussions and new developments on how every aspect of the theory should be understood. This includes the nature of wavefunction/quantum state, the non-commutative structure, the measurement problem \cite{schlosshauer2005decoherence}, entanglement and nonlocality etc. Bell’s theorem\cite{bell1964einstein,bell2004speakable}, one of the pinnacle achievements in QM, was referred to as an “experimental metaphysics”\cite{shimony1984contextual} for its discovery that seemingly metaphysical views about Nature (nonlocality, realism, superdeterminism etc.) can be greatly informed by experimental means. More than that, entanglement and quantum superpositions have practical implications and are now opening up new possibilities in communications, cryptography and computation\cite{nielsen2002quantum}. Even though we have been applying QM for nearly a century now, it is still so deeply mysterious and strange to us human evolved in a classical macroscopic world. No-Go theorems in QM revealed to us exactly how QM is so different from the classical worldview: Kochen-Specker theorem \cite{kochen1975problem} on the noncontextuality of value assignments of quantum observables; Nonlocality theorems on how a causal explanation of quantum correlations must involve nonlocality effects. 

Recently, there has been a new kind of No-Go theorem on the nature of quantum state itself. The Pusey-Barrett-Rudolph theorem (PBR theorem hereafter)\cite{pusey2012reality} and other closely-related \(\psi\)-ontic theorems\cite{leifer2014quantum}, state that in the ontic model framework, quantum states cannot be purely epistemic states representing knowledge about the ontic states. These are No-Go theorems for statistical interpretations of quantum state, hold by Einstein and others who view quantum states as analogous to the role of macrostates and probability distributions in statistical mechanics \cite{ballentine1970statistical,harrigan2010einstein}. 

In 1961, Wigner proposed a thought experiment \cite{wigner1995remarks} (dubbed Wigner’s friend experiment) where he considered the quantum observation of quantum measurement itself. The interesting thought experiment did not make much wave as the discussion was swayed into the direction of the role of observer consciousness on the collapse of quantum state. However, lately there has been some interesting developments in applying Bell-like scenario of quantum correlation measurements to Wigner’s thought experiment involving two Wigner’s friend setups \cite{brukner2017quantum, brukner2018no, frauchiger2018quantum, bong2020strong}.

In this paper, we explore the plausibility of the very idea of ontic states in ontological models within the context of Wigner’s friend setup. We devise two thought experiments to show that ontic states of a physical system must be relative to observer measurement context. The PBR theorem answered the question of whether the same state of reality can be compatible with multiple quantum states. Here we will answer the question of whether the same quantum state can be compatible with multiple states of reality. Just like the answer provided by PBR theorem, our answer is in the negative. More precisely, our first argument shows that these assumptions cannot be all true: ”independent reality of ontic states” (which we call it as “encapsulated measurement non-contextuality), “no-superdeterminism” or the PBR theorem will be violated. Our second argument does away with the “no-superdeterminism” assumption, by making use of an information transfer protocol from Wigner’s friend to Wigner that does not alter the ontic state of the lab. Both of our arguments do not involve correlated observations among two Wigner’s friends setup as in \cite{frauchiger2018quantum, brukner2018no, bong2020strong}.

\section{Encapsulated Measurement and Ontic Model} \label{sec:02}
\textbf{Encapsulated Measurement}: We describe an Encapsulated Measurement (EM) scenario as follows. Observer $F$ performs a measurement $M$ on a quantum system $S$. After receiving a definite outcome, $F$ updates the state of the system to one of the eigenstates of $M$. This measurement process is described as a quantum unitary process by another observer, $W$. To $W$, the combined system $S + F$ is a compound quantum system. Both $F$ and $W$ could be quantum automata, so no issue of consciousness is involved.
Consider the case where $F$ measures the $z$-component of the spin of a spin-\(\frac{1}{2}\) particle $S$ prepared in the state \(\ket{+\hat x} := \ket{\hat S_x =+\frac{1}{2} }\). To observer $F$, upon obtaining the outcome \(\hat S_z = +\frac{1}{2}\) (or \(\hat S_z = -\frac{1}{2}\)), the state of $S$ is updated to \(\ket{+\hat z}\) (to \(\ket{- \hat z}\), respectively). To observer $W$, this process is described by the following unitary evolution:

\begin{align}
    &\ket{+\hat x}_S\ket{F_0}\label{eqn:1} \\ &\xrightarrow{U_t} \ket{\Phi}:=\frac{1}{\sqrt{2}}(\ket{+\hat z}_S\ket{F_+}+\ket{-\hat z}_S\ket{F_-})\label{eqn:2} 
\end{align}
where \(F_0\) is the initial state of $F$; while \(F_+\) and \(F_-\) represent $F$ observing spin-up and spin-down in its $z$-measurement at time $t$.\footnote{The phase between the two terms depends on the Hamiltonian that W uses to model the measurement process, of which he is in control.}

From the perspective of measurement contexts, at $t$, $F$ performs a measurement in the basis \(\{\,\ketbra{+\hat z}{+\hat z},\, I-\ketbra{+\hat z}{+\hat z}\, \} \) on $S$. At the same time, $W$ can justify his state assignment by performing the measurement \(\{\, \ketbra{\Phi}{\Phi},\, I-\ketbra{\Phi}{\Phi}\, \} \) on $S+F$. He will obtain state (\ref{eqn:2}) with probability one. Note that this measurement does not alter the state of $S+F$, and can thus be performed for any number of times. \footnote{Note that we do not assume the observer in the lab, $F$, to assign itself a quantum state. If this is assumed, our conclusion will be much easier to prove.}

This is the encapsulated measurement scenario. We denote this scenario as \(((S)F)W\).\bigskip

\textbf{Ontic Models}: Now we consider the ontic model for \(S+F\). 

For encapsulated measurements, we have to include the ontic state of the observer $F$ into our model. We take this to be the ontic state of the part of the observer that is relevant to the measurement outcome, i.e. the ontic state of $F$ seeing a measurement outcome. To $W$, this is the set of degrees of freedom of $F$ entangled with the observed system during the measurement process. As mentioned above, $F$ could just be any quantum automata that does not involve any consciousness.

Let \(\Omega_{S+F}^+\) be the set of (combined) ontic states of \(S+F\) (i.e. the support), all corresponding to the scenario of $S$ being assigned the state \(\ket{+\hat z}\) by $F$, as $F$ seeing a spin up as outcome. Denote any such state in the set to be \(\lambda_+ \in \Omega_{S+F}^+\).

Our arguments in this paper do not require the use of probability distributions that correspond to quantum states. The concept of support in ontic space is all we required. Support of a quantum state \(\psi\) is the set of ontic states \(\Omega_{\psi}\) where the corresponding distribution of \(\psi\), \(P_{\psi}\) is positive: \(P_{\psi }(\Omega)>0\).

Our arguments invoke PBR theorem in the following form:\bigskip

{\leftskip=15pt\relax
 \rightskip=15pt\relax
\noindent \textit{PBR Theorem:} Different quantum states of a system have non-overlapping supports in the ontic space of the system.\bigskip
\par}

In particular, the PBR theorem will be applied to $S+F$: no ontic state of $S+F$ can be an instance of more than one quantum states in \(\mathcal{H}_{SF}\).

\section{The Arguments} \label{sec:argument}

We provide two different arguments for our result. The first argument makes use of the assumption of superdeterminism. The second argument does not involve superdeterminism, but requires a protocol for information transfer from $F$ to $W$ that does not alter the ontic state of $S+F$.

The flow of both arguments is this: if we assume the ontic state of $S+F$ is the same for $F$ and $W$, i.e. independent of the encapsulated measurement contexts performed by both observers, then the ontic state will be consistent with at least two different quantum states of $S+F$ assigned by $W$. But this is a violation of the PBR theorem (as applied to $S+F$). As a result, there are ontic states of $S+F$ that is different to observers $F$ and $W$, which contradicts the very meaning of an ontic state.

We therefore make the following independence assumption for both of our arguments:\bigskip

{\leftskip=15pt\relax
 \rightskip=15pt\relax
 \noindent \textit{Encapsulated measurement non-contextuality (EMNC)}: 
The ontic state of any system is the same for any observers, including in the encapsulated measurement scenario.\bigskip
 \par}

 Note that this assumption does not say that the ontic state of a system is not altered by different measurements contexts. Rather, it says that the ontic state is the same for different observers that are observing the system at the same time, i.e. the observer and the superobserver. Equivalently, it is the same ontic state for different measurement contexts that are executable at the same time. This is, naturally, a necessary condition of the definition of ontic state - the objectively real state of the system, independent of any observing agents.
 
 We also require another natural assumption that will be used in the arguments. The idea being no-superdeterminism:\bigskip

{\leftskip=15pt\relax
 \rightskip=15pt\relax
\noindent \textit{No superdeterminism:} Choices of observer agent is free in the sense of being independent of earlier states of the world.\bigskip
\par}

Some would say the result is obvious as the superobserver and observer assign different quantum states to the lab: $F$ assigns a product state of $S+F$ corresponding to the observed outcome, while $W$ assigns a superposition of such product states corresponding to all outcomes. Since they are different pure states, their ontic states must be different due to the PBR theorem.

However, for many interpretations of quantum mechanics, the friend cannot assign a quantum state to himself. Such assignment is possible in no-collapse interpretations such as the relative-states and many-worlds interpretation, but then there is only one universal quantum state for the lab, and the $S+F$ is no longer in a product state. Therefore the argument in the previous paragraph cannot proceed. The trick is to produce an argument where the friend does not assign a quantum state to himself. This is what is achieved in the following two arguments. \medskip

\noindent\textbf{Argument (I): Free-Choice Wigner’s Friend Thought Experiment}\label{Arg1}\medskip

In this setup, the lab consists of two qubit systems (spin-half particle) $S$ and $S^\prime$, whose spin components are to be measured by two observers $F$ and $F^\prime$, respectively. Both $S$ and $S^\prime$ are initially prepared in the \(\hat S_x\) eigenstate \(\ket{+\hat x}\) before each trial. Outside the lab, there are two superobservers, lets denote them $W_1$ and $W_2$.

Each trial adheres to the following protocol (see Figure 1):\label{protocol1}\medskip

\noindent At the start of each trial, $F^\prime$ measures the \(\hat{z}\)-component spin of $S^\prime$, i.e. a measurement in the basis \(\{\, \ketbra{+\hat z}{+\hat z}, I - \ketbra{+\hat z}{+\hat z}\, \}\).
\begin{enumerate}
    \item If $F^\prime$ obtains \(\hat S_z = +\frac{1}{2}\), $F$ will similarly measure the \(\hat{z}\)-component spin of $S$.
    \item If $F^\prime$ obtains \(\hat S_z = -\frac{1}{2}\), $F$ will randomly choose to measure one of the two different spin components of $S$, \(\{\hat{n_1}, \hat{n_2}\}\), where \(\hat{n_1}\) and \(\hat{n_2}\) are different directions.
    \item Each measurement trial results can only be accessed by one superobserver, i.e. either by $W_1$ or $W_2$. If the result is one in which $F^\prime$ obtains spin up, then the measurement results in step 1 above will be randomly sent to one of the superobservers. If the result is one in which $F^\prime$ obtains spin down, then those results correspond to $F$ measure direction \(\hat{n_1}\) will be sent to $W_1$, while $W_2$ will receive outcomes corresponding to $F$ measures spin direction \(\hat{n_2}\).
\end{enumerate}
The random choice in step 3 above could be the free choice of either of the superobservers $W_i$, or of a third party free agent located outside the lab that can communicate with $W_1$ and $W_2$.

For each superoberver $W_i$, $i = 1$ or $2$, since he sees $F$ only measures the $z$-direction or the $n_i$-direction, he will model the measurement process for the labs he observe with the Hamiltonian \(exp(-\frac{i}{\hbar}{\hat H} t) = \hat U_{SF} \circ \hat U_{S^\prime F^\prime} \) such that:
\begin{widetext}

\begin{align}
    &\ket{+\hat x}_S\ket{F_0}\ket{+\hat x}_{S^\prime}\ket{F^\prime_0}\label{eqn:3} \\ \xrightarrow{U_{S^\prime F^\prime}} &\ket{+\hat x}_S\ket{F_0}\frac{1}{\sqrt{2}}( \ket{+\hat z}_{S^\prime}\ket{+\hat z}_{F^\prime} + \ket{-\hat z}_{S^\prime}\ket{-\hat z}_{F^\prime} )\label{eqn:4} \\
    \begin{split}
    \xrightarrow{\hspace{2pt}U_{SF}\hspace{2pt}} 
    &\frac{1}{2} ( \ket{+\hat z}_{S}\ket{+\hat z} _{F} + \ket{-\hat z}_{S}\ket{-\hat z} _{F} ) \ket{+\hat z}_{S^\prime}\ket{+\hat z}_{F^\prime} \label{eqn:5} \\
    & +\frac{1}{\sqrt{2}} ( \braket{+\hat n_i}{+ \hat x}_S\ket{+\hat n_i}_S \ket{+\hat n_i}_F + \braket{-\hat n_i}{+ \hat x}_S\ket{-\hat n_i}_S \ket{-\hat n_i}_F ) \ket{-\hat z}_{S^\prime}\ket{-\hat z}_{F^\prime} := \ket{\Psi_{\hat n_i}}
    \end{split}
\end{align}
\end{widetext}
where 
\(\ket{F_0}\) and \(\ket{F^\prime_0}\) are initial states of observers $F$ and $F^\prime$ respectively;
\(\ket{\pm \hat z}_{F^\prime}\) are the states where $F^\prime$ observes a spin up or down in the $z$-component spin measurement of $S^\prime$; \(\ket{\pm \hat z}_F\) are the states where $F$ observes a spin up or down in the $z$-component spin measurement of $S$; \(\ket{\pm \hat n_i}_F\) are the states where $F$ observes a spin up or down in the $n_i$-component spin measurement of $S$.\medskip

Indeed, both superobservers $W_i$ can confirm the state of the lab is \(\ket{\Psi_{\hat n_i}}\) by performing a measurement in the basis \(\{\,\ketbra{\Psi_{\hat n_i}}{\Psi_{\hat n_i}},\, I-\ketbra{\Psi_{\hat n_i}}{\Psi_{\hat n_i}}\, \}\).

\begin{figure}
\includegraphics[width=\columnwidth]{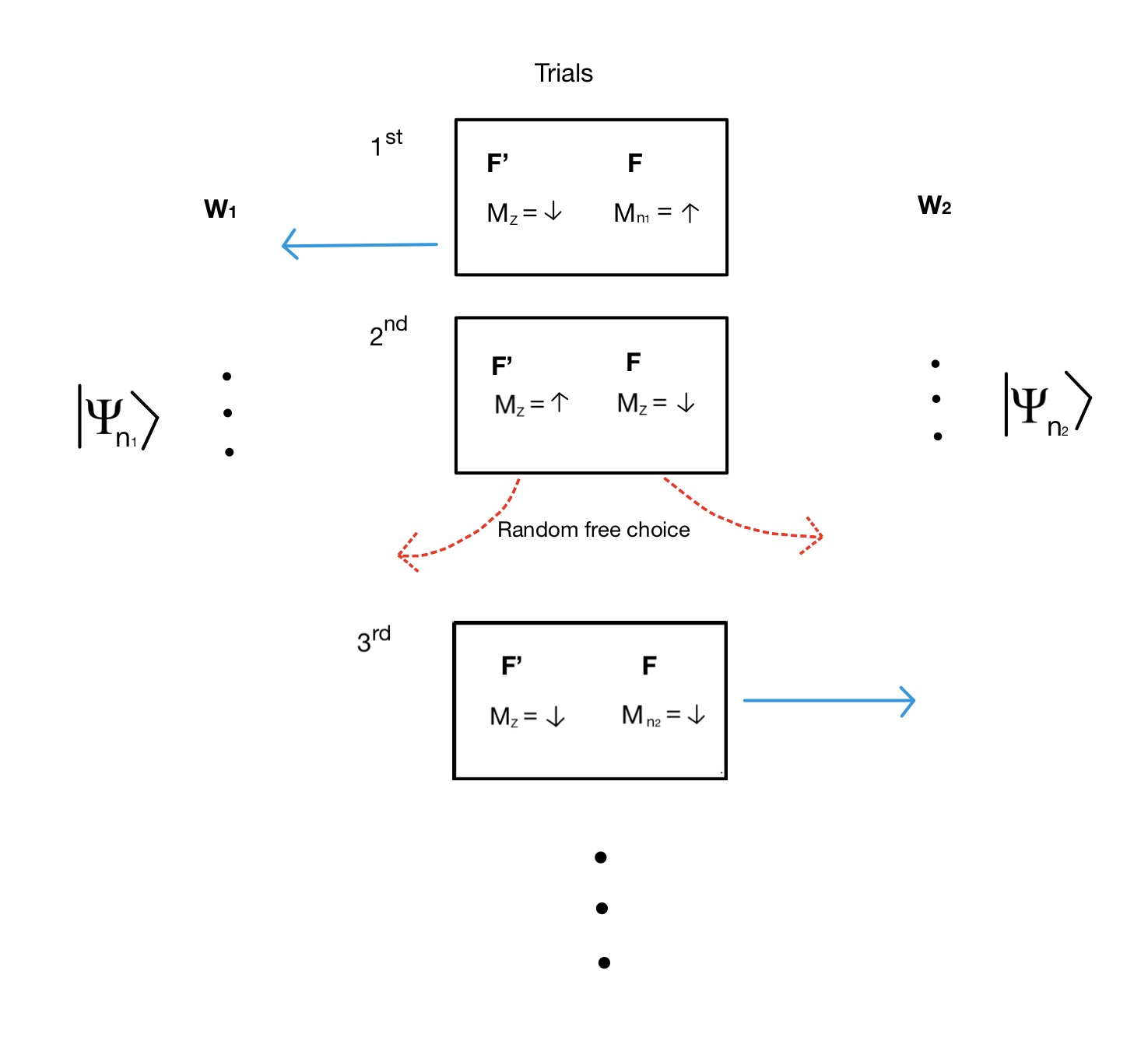}
\caption{A particular sequence of trial outcomes. Each superobserver has access to only part of the lab trials. $W_1$ has access to trials where $F$ measures direction $n_1$, while $W_2$ has access to trials where $F$ measures direction $n_2$. For trials where both friends measure the $z$-direction, it can be chosen freely (by $W_1$ or $W_2$ or a third party free agent) to be accessed by one of the superobservers.}
\centering
\end{figure}

Now, consider a particular sequence of trials like, say, the one shown in Figure 1. Let the ontic state of the lab (\(S + F + S^\prime + F^\prime \)) in the second trial (where both friends measure the same direction $\hat z$) be $\lambda_L$. If this trial happened to be chosen by (or sent by the third party free agent to) superobserver $W_1$, the state of the lab would be \(\ket{\Psi_{\hat n_1}}\) as assigned by $W_1$. Therefore if \textit{EMNC} is true, the ontic state $\lambda_L$ lies within the support of \(\ket{\Psi_{\hat n_1}}\).

But the No-superdeterminism assumption implies that, given the same real ontic state of the lab $\lambda_L$ after the end of the second trial in Figure 1, the lab result may well have been chosen by the other superobserver $W_2$ (or $W_2$ may well have been chosen by the third party free agent to receive that lab result). So the ontic state $\lambda_L$ is also compatible with the state that $W_2$ assigns to the lab, which is \(\ket{\Psi_{\hat n_2}}\).

As a result, the same ontic state lies within the support of two different quantum states that the superobservers (correctly) assign to the lab:  \(\ket{\Psi_{\hat n_1}}\) and  \(\ket{\Psi_{\hat n_2}}\). Thus a contradiction of the PBR theorem as applied to the lab as a whole. 

In summary,\medskip

{\leftskip=15pt\relax
 \rightskip=15pt\relax
\noindent \textit{EMNC + No-superdeterminism + QM \(\implies \neg\)PBR theorem} 
\bigskip
\par}

Taking \textit{No-superdeterminism} to be obviously true \footnote{Although there have been several arguments for backward causation \cite{price2008toy} and superdeterminism\cite{hooft2007free} in QM, we do not think they are viable: the former is paradox-ridden and the latter challenges the validity of Scientific Method itself. But in case No-superdeterminism is a concern, Argument II does away with that assumption.}, and that Quantum Mechanics is correct (so that a superobserver $W$ is correct in assigning a suitable quantum state to the entire lab) and conditions that go into the PBR theorem, we arrive at the conclusion that ontic state of the lab necessarily look different to different observers $F$ and $W$ in the encapsulated measurement scenario.\bigskip

\noindent\textbf{Argument (II): Null-Signal Information transfer protocol}

The scenario in this argument is the standard encapsulated measurement scenario described in the second section - observer $F$ measures the $z$-spin component of the spin-half particle $S$ in the lab, while an outside observer $W$ assigns the state (\ref{eqn:2}) to the lab. 

As in Argument I, there is also a protocol to be followed by both parties in each trial. The aim of this protocol is to allow information transfer from the lab to $W$, so that $W$ updates his state assignment accordingly. According to this protocol, some form of signal will (i.e. with probability one) reach $W$ at time \(t = t_0\). The protocol is as follow:\bigskip
\begin{enumerate}
    \item At time \(t = 0\), $F$ measures the \(z\)-spin component of a qubit $S$ in state \(\ket{+x}_S\).
    \item If $F$ obtains \(\hat S_z = +\frac{1}{2}\), no signal will be sent.
    \item If $F$ obtains \(\hat S_z = -\frac{1}{2}\), the lab will send a signal to $W$, such that $W$ will receive it at time \(t_0\).
\end{enumerate}
It is crucial for our argument that by “no signal will be sent” (a “null signal”), we mean nothing happens in the lab, such that the ontic state of the entire lab remains the same from \(t=0\) to (at least) \(t_0\). Even though no signal will be sent, $W$ knows that 
$F$ observed a spin-up.

According to the protocol, during the time interval \((0, t_0)\), $W$ will assign the following state to the lab:
\begin{align}
\begin{split}
     &\ket{\Psi}:=\\
     &\frac{1}{\sqrt{2}}(\ket{+\hat z}_S\ket{F_+}\ket{0}_L+\ket{-\hat z}_S\ket{F_-}\ket{1}_L)\label{eqn:6}
\end{split}
\end{align}
where \(\ket{0}_L\) denotes the state of the lab where nothing happens, which means no signal is coming out from the lab.

However, if $W$ receives no signal at time \(t_0\), he will have to update the state of the lab to
\begin{align}
    \ket{+\hat z}_S\ket{F_+}\ket{0}_L \label{eqn:7}
\end{align}

\begin{figure}
\includegraphics[width=\columnwidth]{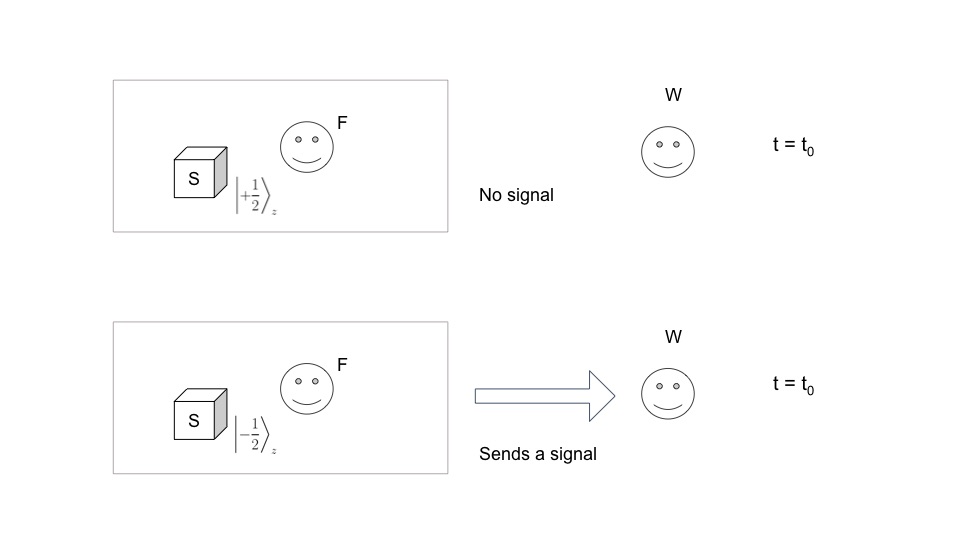}
\caption{ }
\centering
\end{figure}

Now, we consider the scenario where $F$ obtains \(\hat S_z= +\frac{1}{2}\) in her measurement. Let the ontic state of the lab (i.e. $S+F$) right after $F$ receives her measurement outcome at \(t = 0\) be \(\lambda_L(0)\). The \textit{EMNC} assumption implies that the ontic state of the lab will be the same with respect to $W$, which means \(\lambda_L(0)\) lies within the support of the state in \ref{eqn:6}, \(\ket{\Psi}\).

But since $F$ obtains \(\hat S_z = +\frac{1}{2}\), nothing will happen in the lab (resulting in no signal being sent, as required by the protocol). The ontic state of the lab will therefore still be \(\lambda_L(0)\) at \(t_0\): 
\begin{align}
    \lambda_L(t_0) = \lambda_L(0)
\end{align}

As pointed out above, the state of the lab is now \(\ket{+\hat z}_S\ket{F_+}\ket{0}_L\) as assigned by $W$. By the \textit{EMNC} assumption again, the ontic state \(\lambda_L(0)\) must lie within the support of \(\ket{+\hat z}_S\ket{F_+}\ket{0}_L\).

Thus we have shown that the ontic state \(\lambda_L(0)\) lies within the support of two different quantum states of the lab: \(\ket{\Psi}\) and \(\ket{+\hat z}_S\ket{F_+}\ket{0}_L\), a violation of the PBR theorem.

Since the only assumptions required for this argument is \textit{EMNC} and the correctness of QM, to avoid the contradiction we have to abandon the \textit{EMNC} assumption and accept that ontic state of the lab is different with respect to $F$ and $W$.

\section{Discussions}

\noindent \textit{Implications}\medskip

The most straightforward implication of our result would be the ontic states of physical systems (measured systems and observers), if exist, must be relative to observer contexts. That is, the real physical state of the system looks differently to different observers. But this seems to contradict with the very definition of ontic state of a physical system, as it refers to something that objectively exists in the world, independently of any observer. Therefore, the arguments say that independently real physical states do not exist. 

There is another, less obvious possibility. To block the logical chain in our arguments, we may insist that physical systems possess intrinsic ontic states, but reject the possibility that a quantum state corresponds to a set of ontic states. In other words, the same quantum state cannot correspond to multiple states of independent reality. PBR theorem and other \(\psi\)-ontic theorems have shown that one reality cannot correspond to multiple quantum states. Our result, if interpreted this way, implies that the remaining possibility is also not possible, thus denying the very possibility of ontological models for Quantum systems.

Of course, one may deny the assumptions that go into the PBR theorem, such as the the Preparation Independence Postulate (PIP), which, according to Leifer \cite{leifer2014quantum}, is a conjunction of Cartesian Product Assumption (CPA) and No Correlations Assumption (NCA). But the interesting part is, according to the BCLM theorem \cite{barrett2014no}, \(\psi\)-epistemic models where quantum overlaps \(\omega_Q\) are reproduced by overlaps of distributions over ontic space \(\omega_C\), there are two quantum states \(\psi\) and \(\phi\) such that

\begin{align}
    \omega_C (P_{\psi},P_{\phi}) < \frac{2}{d} \, \omega_Q (\ket{\psi},\ket{\phi})
\end{align}

For infinite dimensional Hilbert space (\(d = \infty \)) such as the Hilbert spaces of the labs in our arguments\footnote{We can let the macroscopic observers $F$ and $F^\prime$ to be infinite dimensional quantum systems, as described by $W$.}, quantum state overlaps are as small as it can be. 

However, as shown in our first argument, for the ontic state of the lab in the first trial, we can always pick two quantum states such that the ontic state lies in both of their supports. Those two quantum states could simply be the pair that have overlap that tends to zero, \(\omega_C (P_{\psi},P_{\phi}) \rightarrow 0 \). But this would be a highly unlikely feat according to the BCLM theorem. Since BCLM theorem does not make any of the assumptions invoked in the PBR theorem, we can therefore infer the following:\medskip

\noindent The probability that \textit{EMNC assumption} is wrong is as close to 1 as it can be, even without accepting any of the assumptions in PBR theorems or in other \(\psi\)-ontic theorems.\medskip

In summary, the implications discussed above led to the conclusion that QM does not admit ontic models where quantum states correspond to observer-independent physical states. \bigskip

\noindent \textit{Interpretations of QM}\medskip

It would be interesting to see which interpretations of QM our result lends its support to. \footnote{Physical Collapse theories \cite{Ghirardi2020collapse} are outside the scope of our arguments because the correctness of QM is denied, where $W$ has to assign a collapsed state to the lab, not a superposition of different outcomes.} As mentioned above, the most straightforward implication is to deny the \textit{EMNC} assumption, and accept that the real physical states are relative to observer‘s measurement contexts. Such relational reality seems to support the relational state interpretation (RQM) of Rovelli \cite{laudisa2002relational} or even the original relative-state interpretation of Everett \cite{barrett2018everett}, but this is not true. What is relational or relative in our argument is the physically real state of the system, with respect to measurement contexts in an encapsulated measurement scenario. Rovelli’s RQM is about the physical variables of a system having definite values relative to another system with which it is entangled with. Everett's relative-state interpretation is about the quantum state of a system being relative to the quantum state of an observer, with whom it is in an entanglement. Unlike our result, none of these interpretations is about the relativity between the physically real state and a measurement context.

The neo-Copenhagen interpretations are in general consistent with results obtained in this paper. For example, in QBsim‘s participatory realism\cite{fuchs2017participatory,fuchs2017notwithstanding}, different measurement contexts create different physical realities. This is another way of stating the relative aspect of physical state with respect to measurement contexts.\footnote{Actually, QBism goes further than that by stating that (certain) physical reality is subjective, i.e. may be distinct for different observers, not only to distinct contexts. The results in this paper do not go that far.} Another possible option is a neo-Copenhagen interpretation that develops along the Bohrian viewpoint \footnote{Paper is forthcoming by this author.}, where measurement outcomes have a context-dependent definiteness. Physical state of the lab becomes context dependent if we include the outcomes as part of the physical reality.

\bigskip

\bibliography{mybibliography}

\end{document}